\documentclass[aps,prl,showpacs,amsmath,twocolumn,amssymb,superscriptaddress,letterpaper]{revtex4}
\usepackage{mathrsfs}
\usepackage{graphicx}
\usepackage{graphics}
\usepackage{subfigure}
\usepackage{bm}
\usepackage{dcolumn}
\usepackage{amsmath,bm}
\usepackage{color}

\begin{document}
\title{Disorder and metal-insulator transitions in Weyl semimetals}

\author{Chui-Zhen Chen}
\affiliation{International Center for Quantum Materials, School of Physics,
Peking University, Beijing 100871, China}
\affiliation{Collaborative Innovation Center of Quantum Matter, Beijing, 100871, China}
\author{Juntao Song}
\affiliation{Department of Physics, Hebei Normal University, Hebei 050024, China}
\author{Hua Jiang}
\affiliation{College of Physics, Optoelectronics and Energy,
Soochow University, Suzhou 215006, China}
\author{Qing-feng Sun}
\affiliation{International Center for Quantum Materials, School of Physics,
Peking University, Beijing 100871, China}
\affiliation{Collaborative Innovation Center of Quantum Matter, Beijing, 100871, China}
\author{Ziqiang Wang}
\affiliation{Department of Physics, Boston College, Chestnut Hill, Massachusetts 02167, USA}

\author{X. C. Xie}
\affiliation{International Center for Quantum Materials, School of Physics,
Peking University, Beijing 100871, China}
\affiliation{Collaborative Innovation Center of Quantum Matter, Beijing, 100871, China}
\date{\today}

\begin{abstract}
The Weyl semimetal (WSM) is a newly proposed quantum state of matter. It
has Weyl nodes in bulk excitations and Fermi arcs surface states. We study the effects of disorder and localization in WSMs and find three exotic phase transitions.
(I) Two Weyl nodes near the Brillouin zone boundary can be annihilated pairwise by disorder scattering, resulting in the opening of a topologically nontrivial gap and a transition from a WSM to a three-dimensional (3D) quantum anomalous Hall state. (II) When the two Weyl nodes are well separated in momentum space, the emergent bulk extended states can give rise to a direct transition from a WSM to a 3D diffusive anomalous Hall metal.
(III) Two Weyl nodes can emerge near the zone center when an insulating gap closes with increasing disorder, enabling a direct transition from a normal band insulator to a WSM.
We determine the phase diagram by numerically computing the localization length and the Hall conductivity,
and propose that the exotic phase transitions can be realized on a photonic lattice.
\end{abstract}


\maketitle


Topological quantum states of matter have emerged as an important and growing field in condensed matter and materials physics recently \cite{Hasan2010,Qi2011}. The Weyl semimetal (WSM) is a newly proposed quantum state of the kind that breaks time-reversal symmetry or inversion symmetry \cite{Wan2010,Yang2011,Xu2011,Burkov2011,Weng2015,Lv2015,Xu2015,Lu2015}.
A WSM exhibits a set of paired zero-energy Weyl nodes (linearly touching points of conduction and valence bands) in its bulk spectrum and Fermi arcs excitations localized on the surface.
A number of candidate materials have been predicted to be WSMs, including pyrochlore iridates and magnetic topological insulator multilayers \cite{Wan2010,Yang2011,Xu2011,Burkov2011}.
Recently, following the theoretical prediction \cite{Weng2015}, angle resolved photoemission experiments confirmed that TaAs is a WSM by the observation of the Fermi arcs surface states \cite{Lv2015}. Both the Weyl nodes and the Fermi arcs have been observed in NbAs using a combination of soft X-ray and ultraviolet photoemission experiments \cite{Xu2015}. Furthermore, the Weyl points have been predicted and subsequently observed remarkably in gyroid photonic crystals \cite{Lu2015}.

In this Letter, we study both numerically and analytically the stability of the gapless Weyl nodes and Fermi arcs against random potential scattering and the novel disorder-induced metal-insulator transitions in WSM systems. Previous studies have concentrated on the properties of a single Weyl node, assuming that the disorder potential is smooth enough to avoid scattering between different nodes \cite{Hosur2012,Ominato2014,Biswas2014,Sbierski2014}. Indeed, a system with a single Weyl node is not subject to Anderson localization even for strong disorder \cite{Ryu2010}.
However, the theorem of Nielsen and Ninomiya states that
gapless Weyl nodes with opposite chirality must appear in pairs
\cite{Nielsen1981}. Thus, it is essential to study the localization properties of a pair of Weyl nodes since they can be annihilated pairwise when approaching each other in momentum space or by strong intervalley scattering \cite{Fradkin1986,Lu_HZ2015}. To this end, we study a model system of quasi-3D WSMs formed by stacking 2D quantum anomalous Hall (QAH) layers along the $z$-direction \cite{Burkov2011} which can support a single-pair of Weyl nodes in the first Brillouin zone (BZ), and uncover a rich set of novel disorder and localization effects. Consider the case when the Weyl nodes are located close to each other, which can happen either near the zone boundary or the zone center. We find that in the former case, the Weyl points are unstable against disorder, which causes the pair to merge at the zone boundary and annihilate each other, whereby opening up a topologically nontrivial gap in the bulk. This gives rise to a transition from the WSM to the 3D QAH state with fully quantized Hall conductivity. In the latter case, we find that the critical state with two overlapping Weyl nodes at the zone center is stable against disorder as they would repel each other and lift the degeneracy in favor of a WSM. Moreover, if one starts with a normal insulator with a band gap at the zone center, increasing disorder would lead to the closing of the band gap and the emergences of two repelling Weyl nodes near the zone center, i.e. a normal band insulator to WSM transition. It is thus rather remarkable that tuning the disorder strength can switch the system between a WSM and a normal band insulator, as well as between a 3D QAH and a WSM, in analogy to the``phase switching'' between a normal band insulator and a topological insulator by disorder \cite{Jiang2009,Li2009,Groth2009,Titum2015}. A pair of Weyl nodes well separated in momentum space are found to be stable against weak disorder. However, we find that strong enough disorder induces bulk extended states and gives rise to a direct transition from the WSM to a
3D diffusive anomalous Hall metal \cite{Kobayashi2014,Sbierski2014}. This disorder induced transition is unconventional since it takes place between two metallic states and is only enabled by the topological character of the WSM.
%
We study the various multiple disorder-driven phase transitions and obtain the phase diagram by evaluating the localization length and the Hall conductivity
\cite{MacKinnon1981,MacKinnon1983,MacKinnon1993,Prodan2013,Prodan2011,Song2014}.
We also performed calculations using the self-consistent Born approximation (SCBA) in order to gain important analytical insights into the stability of the Weyl nodes at weak disorder.

\begin{figure}[bht]
\centering
\includegraphics[scale=0.4, bb = 10 0 600 350, clip=true]{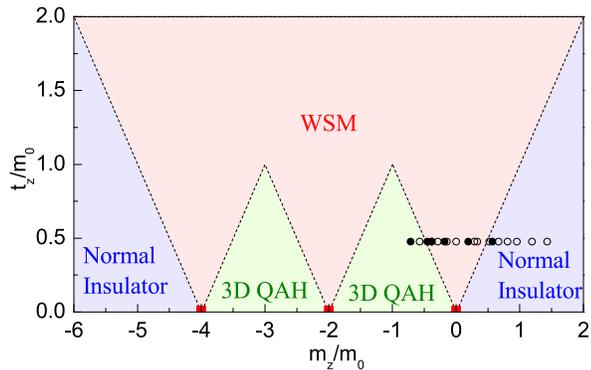}
\caption{(Color online). The phase diagram of the Weyl semimetal
Hamiltonian $H_0$ on the $t_{z}/m_{0}$-$m_{z}/m_{0}$ plane. The open and filled black circles correspond to the parameter values studied for the disordered phase diagram shown in Fig.\ref{fig2}; and the filled black circles to the localization length and Hall conductivity shown in Figs.\ref{fig3} and \ref{fig4}. \label{fig1} }
\end{figure}
We begin with the simple 2$\times$2 WSM Hamiltonian written in momentum space \cite{Yang2011}
\begin{eqnarray}
  H_{0} &=& (m_{z}-t_{z}\cos k_{z} ) \sigma_{z} + m_{0}(2-\cos k_{x}-\cos k_{y})\sigma_{z}  \nonumber \\
  &&   + t_{x}\sigma_{x} \sin k_{x} +t_{y}\sigma_{y} \sin k_{y},
  \label{equ:1}
\end{eqnarray}
with the model parameters $t_{x,y,z}$, $m_{z}$, $m_{0}$ and Pauli matrices $\sigma_{x,y,z}$.
The lattice wave vectors $k_{x,y,z}$ are defined in the first BZ of a $L_x\times L_y\times L_z$ cubic lattice with the lattice constant $a\equiv1$.
$H_{0}$ can be regarded as describing a layered (labeled by $k_z$) Dirac system coupled along the $z$-direction by the inter-layer coupling $t_z$. Diagonalizing $H_0$ gives the energy-momentum relation
\begin{equation}
E_k=\pm\sqrt{\Delta_z^2+(t_x\sin k_x)^2+(t_y\sin k_y)^2}
\label{dispersion}
\end{equation}
where $\Delta_z=m_z-t_z\cos k_z+m_0(2-\cos k_x-\cos k_y)$. Note that $H_0$ breaks the time reversal symmetry when the first two terms (proportional to $\sigma_z$) in Eq.(\ref{equ:1}) are nonzero, i.e. when $\Delta_z\ne0$. Adding the random on-site potential to $H_{0}$, we obtain the full Hamiltonian with disorder
\begin{equation}
H= H_{0} + \left(
              \begin{array}{cc}
                V_{1}({\bf r}) & 0 \\
                0 & V_{2}({\bf r}) \\
              \end{array}
            \right)
            \label{h}
\end{equation}
where  $V_{1,2}({\bf r})$ are uniformly distributed within $[-W/2$,$W/2]$ with $W$ representing the disorder strength.

The phase diagram of the clean WSM Hamiltonian $H_0$ is shown in Fig.\ref{fig1}. When $t_{z}=0$, the layers decouple. There are three critical points on the phase diagram denoted by the red solid squares with massless Dirac nodes located at the 2D BZ center $(0,0)$ for $m_z/m_0=0$ and at the zone boundaries $(0,\pi)$ and $(\pi,0)$ for $m_z/m_0=-2$ and $(\pi,\pi)$ for $m_z/m_0=-4$. They separate three gapped phases: the normal insulator (NI) and two QAH phases with 2D Hall conductances $\sigma_{xy}^{2D}=-e^2/h$ (left) and $e^2/h$ (right) respectively \cite{Qi2006}. Turning on the inter-layer coupling $t_z$ naturally leads to two 3D QAH phases with a bulk topological gap and fully quantized Hall {\em conductivity} $\sigma_{xy}=\pm e^2/h$ \cite{Burkov2011}.
More importantly, Fig.\ref{fig1} shows that the WSM phase emerges between the 3D QAH states and the NIs and is bordered by a pair of phase boundaries emanating from each of the three 2D critical points described by, from left to right, $m_z=\pm t_z-4m_0$, $\pm t_z -2m_0$, and $\pm t_z$. The WSM phase supports pairs of gapless Weyl nodes split by $t_z$ along the $k_z$-direction and are located at $[\pi,\pi,\pm\arccos (m_z+4m_00)/t_z]$; $[0,\pi,\pm\arccos(m_z+2m_0)/t_z]$ and $[\pi,0,\pm\arccos(m_z+2m_0)/t_z]$; and $[0,0,\pm\arccos m_z/t_z]$ between the corresponding pairs of phase boundaries respectively.

\begin{figure}
\centering
\includegraphics[scale=0.35, bb = 50 35 800 525, clip=true]{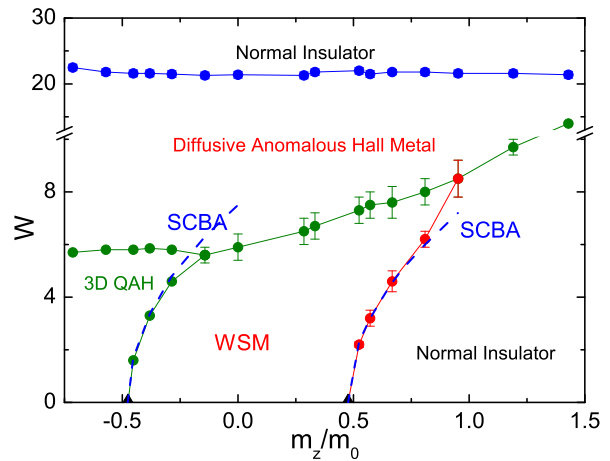}
\caption{(Color online). The phase diagram on the $W-m_{z}$ plane. The symbols guided by the solid lines are obtained from the localization length. The two solid triangles correspond to the phase boundaries in the clean limit.
The blue dashed lines are the phase boundaries determined using the SCBA.
\label{fig2} }
\end{figure}
In the rest of the paper, we will, for the sake of simplicity, focus on the parameter region $-0.714 < m_z/m_0 < 1.429$ in the clean phase diagram (Fig.~\ref{fig1}) by setting $t_{x,y,z}/m_0=0.476$ and investigate the disorder induced phase behaviors with varying disorder strength $W$. The WSM phase in this regime $(-1\le m_z/t_z\le1)$ has a single pair of Weyl nodes located at $(0,0,\pm k_z^0)$, $k_z^0=\arccos (m_z/t_z)$. They move from the zone center $(0,0,0)$ to the zone boundary $(0,0,\pi)$ as the WSM traverses the NI and the 3D QAH phase boundaries.

Our main results are summarized in the phase diagram in the presence of disorder, which is shown in Fig.\ref{fig2} on the $W-m_{z}$ plane. There are four distinct phases and five possible multiple phase transitions with increasing disorder strength $W$: (i) QAH--metal--NI, (ii) WSM--QAH--metal--NI, (iii) WSM--metal--NI, (iv) NI--WSM--metal--NI, and (v) NI--metal--NI. The phase boundaries indicated by the symbols and guided by the solid lines are obtained by numerical computations of the localization length and the Hall conductivity. Note that the WSM--QAH and WSM--NI phase boundaries at weak disorder agree remarkably well with the ones obtained analytically using the SCBA described in the supplemental section. The findings of such weak disorder induced transitions between electronic states of different topological characters broaden the concept of topological Anderson insulators \cite{Jiang2009,Li2009,Groth2009,Titum2015} and support the generality of the rich physics behind the interplay of randomness and topology.
%

\begin{figure}[thb]
\centering
\includegraphics[scale=0.3, bb = 0 0 1000 550, clip=true]{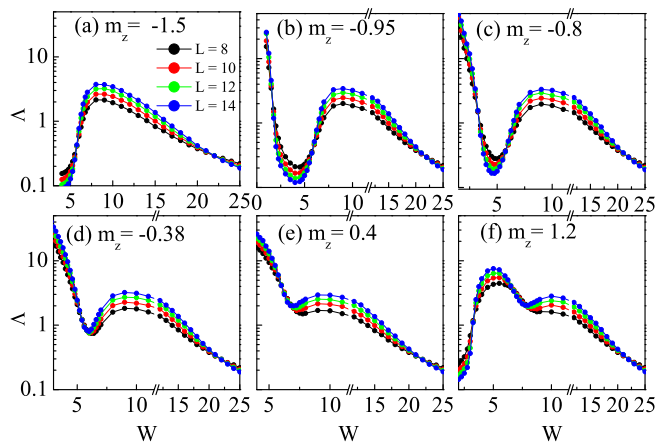}
\caption{(Color online). (a)-(f) Normalized localization length  $\Lambda=\lambda(L)/L$ versus disorder strength $W$ at different mass $m_z$. The curves correspond to different sample widths $L=8,10,12,14$. Other parameters are $m_0=2.1$ and $t_{x,y,z}=1$.
\label{fig3} }
\end{figure}

To calculate the localization length, we consider a 3D long bar sample of length $L_z$ and widths $L_x=L_y=L$ with periodic boundary conditions in the $x$ and $y$-directions. The localization length $\lambda(L)$ of the sample is calculated using the transfer matrix method \cite{MacKinnon1981,MacKinnon1983,MacKinnon1993}. In general, the {\em normalized} localization length $\Lambda\equiv\lambda(L)/L$ increases with $L$ in a metallic phase, decreases with $L$ in an insulating phase, and is independent of $L$ at the critical point of the phase transition. The scale-dependent behaviors of $\Lambda$ versus the disorder strength $W$, obtained at different values of $m_z$ and shown in Fig.~\ref{fig3}, reveal a sequence of disorder induced Anderson transitions in our model.
In Fig.\ref{fig3}(a), the decrease of $\Lambda$ with increasing $L$ at small $W$ places the system at $m_z/m_0=-0.714$ in the 3D QAH state at weak disorder on the phase diagram (Fig.\ref{fig2}). Increasing $W$ encounters a critical point at $W\simeq 6.0$ where $d\Lambda/dL=0$, beyond which a diffusive metallic phase emerges as indicated by $d\Lambda/dL >0$. The latter has a nonzero Hall conductivity (see below) and is thus identified with an anomalous Hall (AH) metal. Further increasing $W$ causes another metal-insulator transition in Fig.\ref{fig3}(a) as the system becomes an Anderson insulator at strong disorder.
Such disorder-induced multiple phase transitions (QAH--AH metal--NI) are analogous to those found in multilayer quantum Hall structures and 3D $Z_{2}$ topological insulators \cite{Chalker1995,Wang1997,Kobayashi2013}.

Repeating this procedure at different values of $m_z/m_0$ maps out the disordered phase diagram shown in Fig.\ref{fig2}. Note that the phase boundary between the QAH and WSM is slanted toward the WSM side, which indicates that a clean WSM in proximity to the QAH state (i.e. when the Weyl nodes are close to $k_z^0=\pi$) would be unstable against disorder $W$ and would undergo a sequence of WSM--QAH--metal--NI transitions with increasing $W$. This unexpected behavior can be deduced directly from the scale-dependence of the localization length in Figs.\ref{fig3}(b)-(c) obtained at $m_{z}/m_0=-0.452$ and $-0.381$. To understand this result, we performed SCBA valid for weak disorder analytically in the supplemental section \cite{Supp}, which amounts to determine the band renormalization induced by disorder \cite{Jiang2009,Li2009,Groth2009,Titum2015}. Interestingly, the most important disorder-induced renormalization in SCBA is for the topological mass $m_z$, which enhances the band inversion and causes the Weyl nodes to ``attract'' each other to the zone boundary. When the they meet at the zone boundary, the Weyl nodes annihilate pairwise and open up a nontrivial bulk gap, leading to the transition from the WSM phase to the 3D QAH phase. This phase transition can also be characterized by the disappears of the Fermi arcs in favor of fully quantized Hall conductivity to be discussed later. One can apply this analysis to the phase boundary between the WSM and the NI in Fig.\ref{fig2} and find another surprising result. The WSM in proximity to the NI with the pair of Weyl nodes close to the zone center is stable against weak disorder since the effect of renormalizing $m_z$ by disorder will cause the Weyl nodes to repel and move apart. This explains why the phase boundary is slanted toward the NI side in Fig.\ref{fig2} and thus an unexpected transition from a NI to a WSM with increasing disorder. Physically, as the insulating band gap closes, a pair gapless Weyl nodes is nucleated at the zone center. The scale-dependence of $\Lambda$ shown in Fig.\ref{fig3}(f) at $m_z/m_0=0.571$ confirms this with the switching from $d\Lambda/dL<0$ to $d\Lambda/dL>0$ at a critical $W\approx3.3$. Thus, a pair of Weyl nodes with opposite chirality are stable against intervalley scattering at weak disorder, provided that they are sufficiently away from the zone boundary.
Remarkably, for strong enough disorder, we find a direct transition from the WSM to the AH metal that enables the WSM--metal--NI and NI--WSM--metal--NI transitions shown in the phase diagram. Such a phase transition comes from the emergent bulk extended states induced by strong disorder \cite{Kobayashi2014,Sbierski2014} and is thus an unusual transition between two metallic states as can be seen from the positive scale-dependence $d\Lambda/dL>0$ on both sides of the critical point in Figs.\ref{fig3}(d)-(e), and on the two sides of the second scale-invariant point in Fig.\ref{fig3}(f). Indeed, the WSM and the 3D diffusive AH metal can be distinguished by the nature of the Hall conductivity quantization.

\begin{figure}
\centering
\includegraphics[scale=0.35, bb =0 0 1000 550, clip=true]{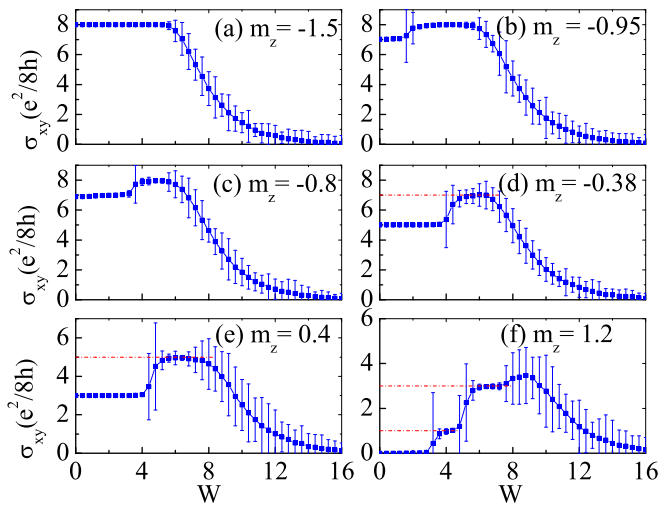}
\caption{(Color online). (a)-(f) Ensemble averaged Hall conductivity
$\sigma_{xy}$ as a function of the disorder strength $W$ on $30 \times 30  \times 8$ samples. The parameters are from $m_{z}=-1.5$ to $m_{z}=1.2$ in one to one correspondence to Fig.\ref{fig3}(a)-(f).
The red dash-dotted lines are guide to the eye. The error bars are magnified ten times to show conductance fluctuations. \label{fig4}
}
\end{figure}
To study the Hall conductivity and the Fermi arcs in WSMs, we view the Hamiltonian $H_{0}$ as a set of 2D massive Dirac Hamiltonian with fixed $k_{z}$ as in Burkov et al. \cite{Burkov2011}. The 2D quantized Hall conductance in the $x-y$ plane at a fixed $k_{z}$ is given by $\sigma_{xy}^{2D}(k_{z})=\Theta(k_z^0-|k_{z}|)e^2/h$
\cite{Burkov2011,Xu2011,Qi2006,Jiang2012}. The total Hall conductivity of the system is $\sigma_{xy}=\sum_{k_{z} \in BZ} \sigma^{2D}_{xy}/L_{z}$ where $k_{z}$ takes on quantized values in the first BZ of $L_{z}$ layers under the periodic boundary condition. Thus $\sigma_{xy}$ is proportional to $2k_z^0$, the distance between the two Weyl nodes. It increases from zero to $e^2/h$ as the pair of Weyl nodes moves from $k_z^0=0$ at the WSM--NI boundary to $k_z^0=\pm\pi$ at the WSM--3D QAH boundary. As a result, the Hall conductivity can characterize the positions of the Weyl nodes as well as the density of Fermi arcs surface states.

In the presence of disorder, we calculate the ensemble averaged Hall
conductivity $\sigma_{xy}$  of quasi-3D samples with $L_{x} \times L_{y} \times L_{z}  = 30 \times 30  \times 8$ using the
noncommutative Kubo formula under periodic boundary conditions in
all three directions \cite{Prodan2013,Prodan2011,Song2014,footnote}.
The results are shown in Figs.\ref{fig4}(a)-(f) at the corresponding parameters of the localization lengths in Figs.\ref{fig3}(a)-(f). Overall, $\sigma_{xy}$ increases with the disorder strength $W$ initially and then decreases with $W$ in the strong disorder limit. Fig.\ref{fig4}(a) shows that the 3D QAH phase sustains a fully quantized conductivity $\sigma_{xy} = e^2/h$ until the system enters the diffusive AH metal phase at $W\simeq6.0$ where bulk extended states emerge and $\sigma_{xy}$ looses quantization, decreases with $W$, and exhibits conductance fluctuations, in complete agreement with the phase diagram obtained from the localization length shown in Fig.\ref{fig3}(a). Figs.\ref{fig4}(b)-(c) start in the WSM phase close to the QAH phase boundary and thus show the ``fractional'' quantized $\sigma_{xy}=7e^2/8h$. Upon increasing $W$, the pairs of Weyl nodes approaches the zone boundary and annihilate each other at a critical $W$, while the Fermi arcs turn into a set of chiral edge states and the Hall conductivity acquires the maximum quantized value $e^2/h$ of the QAH state \cite{footnote2}.
Further increasing $W$ leads to the sequence of QAH--metal--NI transitions.
Interestingly, Figs.\ref{fig4}(d)-(e) show that when the Weyl nodes are sufficiently away from the zone boundary, increasing disorder leads to the discretized growth of the distance between the Weyl points and thus the increase in the fractions of the quantized Hall conductivity. However, the Weyl nodes do not reach the zone boundary before being destroyed by disorder amid a direct transition from WSM to 3D diffusive AH metal as the Fermi arcs scatter through the emergent bulk extended states \cite{Kobayashi2014,Sbierski2014}. The Hall conductivity is no longer quantized. It decreases with disorder and vanishes at the transition to the Anderson insulator at strong disorder. Thus, a Weyl semimetal can be distinguished from a diffusive anomalous Hall metal by the quantized fractions of the Hall conductivity. Finally, Fig.\ref{fig4}(f) shows that as the band gap closes with increasing disorder, a NI with $\sigma_{xy}=0$ can make a transition to a WSM at a critical $W\simeq3.3$, by nucleating a pair of Weyl nodes at the zone center such that $\sigma_{xy}$ increases from $0$ to $e^2/8h$ and then to $3 e^2/8h$ as the distance between the nodes and the density of the Fermi arcs states increase with disorder. At $W\simeq8$ and before the Weyl nodes can reach the zone boundary, the WSM--AH metal transition takes place. We stress that all the phase behaviors obtained from the Hall conductivity are in quantitative agreement with those determined from the localization length in Fig.\ref{fig3} and displayed in the phase diagram in Fig.\ref{fig2}, including the location of the phase boundary. This further demonstrate the consistency and reliability of the obtained results.

In summary, we studied the effects of disorder and Anderson transition in the simplest WSMs with two Weyl nodes and obtained an unexpectedly rich phase diagram. We find that weak disorder has important effects when the Weyl nodes are close in momentum space. The pair-annihilation near the zone boundary and the pair-nucleation at the zone center leads to WSM to QAH as well as NI to WSM transitions. Moderately strong disorder, on the other hand, produces a WSM to diffusive AH metal transition. These results have important implications for real WSM materials discovered recently \cite{Weng2015,Lv2015,Xu2015} which have multiple pairs of Weyl nodes. Furthermore, the WSM phase has been observed recently in gyroid photonic crystals where disorder can be introduced and controlled by a speckled beam \cite{Lu2015,OpL2007}. In photonic crystals, the bulk gap can be detected by bulk transmission measurements while the highly directional surface transmission spectroscopy can be used to detect the chiral surface state through its unidirectional group velocity \cite{Lu2014}. We therefore propose that
the exotic phase transitions found here can be readily tested on optical lattices such as in gyroid photonic crystals.

{\em Acknowledgements.}
We thank Haiwen Liu for fruitful discussions.
This work was financially supported by NBRP
of China (2012CB821402, 2015CB921102, and 2012CB921303) and
NSF-China under Grants Nos.11274364.
ZW is supported by DOE
Basic Energy Sciences grant DE-FG02-99ER45747.
HJ is supported by the NSF of Jiangsu province BK20130283.


\begin{thebibliography}{Weyl}
\bibitem{footnote}The boundary condition in the z-direction will not sigificant influence our results in thermodynamic limit.
\end{thebibliography}
\end{document}


\title{Supplementary materials for ``Disorder and metal-insulator transitions in Weyl semimetals''}
\author{Chui-Zhen Chen}
\affiliation{International Center for Quantum Materials, School of Physics,
Peking University, Beijing 100871, China}
\affiliation{Collaborative Innovation Center of Quantum Matter, Beijing, 100871, China}
\author{Juntao Song}
\affiliation{Department of Physics, Hebei Normal University, Hebei 050024, China}
\author{Hua Jiang}
\affiliation{College of Physics, Optoelectronics and Energy,
Soochow University, Suzhou 215006, China}
\author{Qing-feng Sun}
\affiliation{International Center for Quantum Materials, School of Physics,
Peking University, Beijing 100871, China}
\affiliation{Collaborative Innovation Center of Quantum Matter, Beijing, 100871, China}
\author{Ziqiang Wang}
\affiliation{Department of Physics, Boston College, Chestnut Hill, Massachusetts 02167, USA}

\author{X. C. Xie}
\affiliation{International Center for Quantum Materials, School of Physics,
Peking University, Beijing 100871, China}
\affiliation{Collaborative Innovation Center of Quantum Matter, Beijing, 100871, China}
\date{\today}
\maketitle

\section{self-consistence Born approximation (SCBA)}
In this section, we provide some details of the SCBA calculation. We follow the discussions of Groth et. al. \cite{Groth2009} who performed SCBA calculations for the weak disorder induced transition from a normal band insulator to a topological insulator in systems with time reversal symmetry. We start with
the clean part of the Hamiltonian for WSMs, i.e. Eq.(1) in the main text,
\begin{eqnarray}
  H_{0} &=& (m_{z}-t_{z}\cos k_{z} ) \sigma_{z} + m_{0}(2-\cos k_{x}-\cos k_{y})\sigma_{z}
     + t_{x}\sigma_{x} \sin k_{x} +t_{y}\sigma_{y} \sin k_{y},
  \label{equ:0}
\end{eqnarray}
where the model parameters $t_{x,y,z}$, $m_{z}$, $m_{0}$ have the same meanings as in the main text. Adding the random on-site potential to $H_{0}$ gives the full Hamiltonian in the presence of disorder
\begin{equation}
H= H_{0} + \left(
              \begin{array}{cc}
                V_{1}({\bf r}) & 0 \\
                0 & V_{2}({\bf r}) \\
              \end{array}
            \right)
            \label{h}
\end{equation}
where  $V_{1,2}({\bf r})$ are uniformly distributed within $[-W/2$,$W/2]$ with $W$ representing the disorder strength.
\begin{figure}[bht]
\centering
\includegraphics[scale=0.4, bb = 0 30 700 530, clip=true]{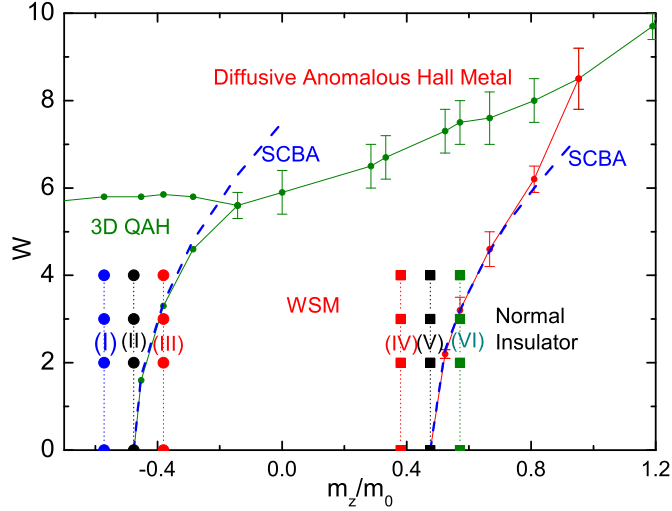}
\caption{(Color online). The phase diagram on the $W-m_{z}$ plane reproduced partly from Fig.2 in the main text. The scattered symbols guided by the solid lines are obtained from finite size scaling of the localization length. The blue dashed lines are the phase boundaries obtained using the SCBA.
The solid dots and squares connected by the vertical dashed lines are the cuts where the SCBA renormalized band dispersions are shown in Fig.S~\ref{Fig1S}. They are labeled by the roman numerals (I) to (VI) in one to one correspondence to the six vertical panels in Fig.S~\ref{Fig1S}.
\label{Fig0S}}
\end{figure}

Within the single-particle picture, the effects of disorder can be accounted for in terms of the self-energy defined as
\begin{equation}
{1\over E_{F}-H_{0}-\Sigma(E_{F})}=\langle {1\over E_{F}-H}\rangle,
\label{selfenergy}
\end{equation}
where $\langle ... \rangle $ denotes the average over disorder configurations. It is convenient to expand the $2\times2$ self-energy matrix in the basis of Pauli matrices
\begin{equation}
\Sigma(E_{F})=\Sigma_{0}\sigma_{0}+\Sigma_{x}\sigma_{x}+\Sigma_{y}\sigma_{y}
+\Sigma_{z}\sigma_{z},
\end{equation}
which allows us to identify the renormalized topological mass and the chemical potential from the self-energy corrections,
\begin{eqnarray}
  \bar{m}_{z} &=&  m_{z} + \lim_{k\rightarrow 0} Re\Sigma_{z},\nonumber\\
  \bar{E}_{F} &=& E_{F} + \lim_{k\rightarrow 0}  Re\Sigma_{0}. \label{equ:1}
\end{eqnarray}
\begin{figure}[bht]
\centering
\includegraphics[scale=0.6, bb = 0 130 1000 650, clip=true]{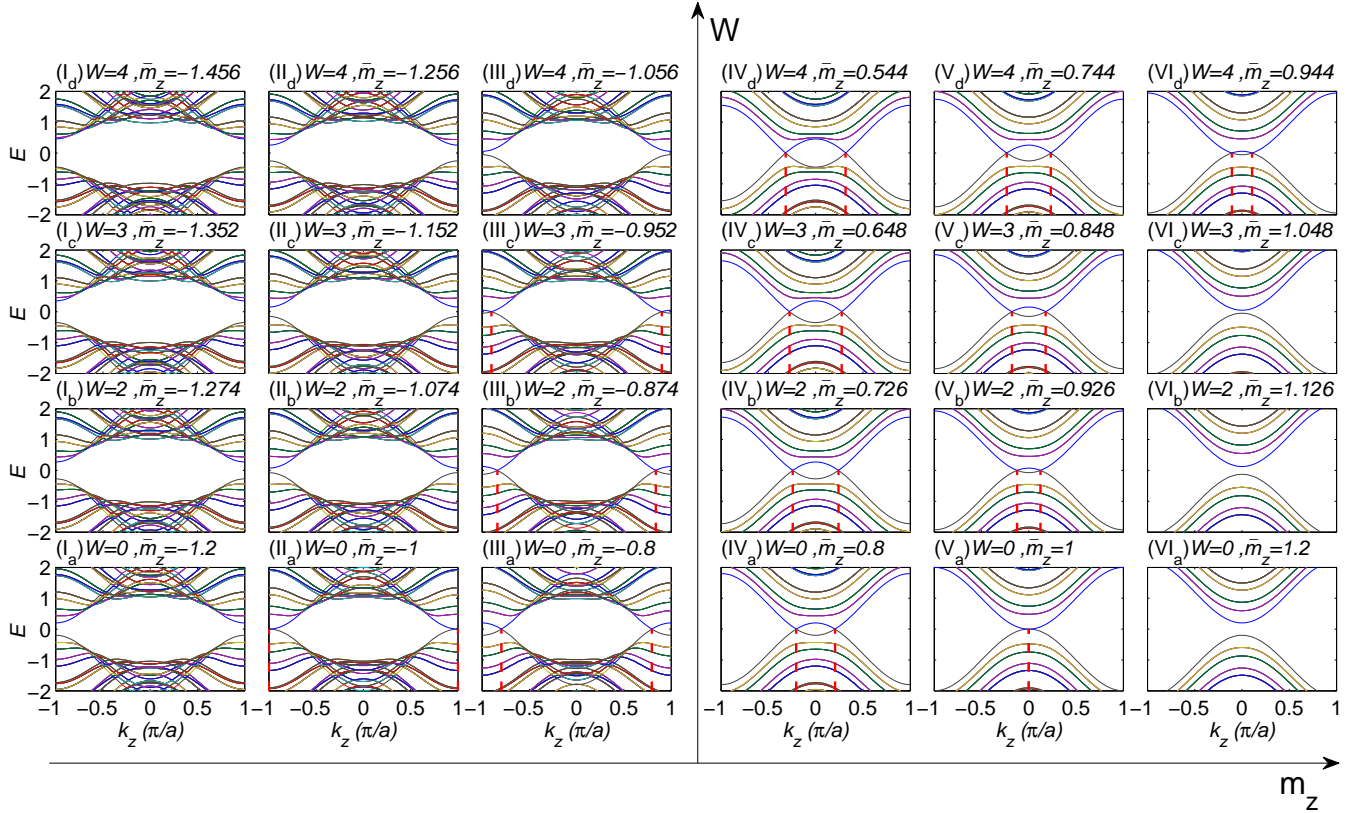}
\caption{(Color online). Panels (I)-(VI) show the evolution of the SCBA dispersion  with increasing disorder strength $W$ (moving up vertically) and increasing mass $m_z$ (moving to the right horizontally). The dispersion is calculated from a 3D long bar sample of length $L_z=\infty$ along the $z$-direction and widths $L_x=L_y=14a$ under periodic boundary conditions in the $x$ and $y$ directions. The SCBA renormalized mass $\bar{m}_{z}$ is obtained by solving Eq.~\ref{equ:1} numerically.
The red dashed lines are guide to the eye for the positions of the Weyl nodes.
The parameters in (I)-(VI) are in one to one correspondence to the six cuts labeled by (I)-(VI) in Fig.S~\ref{Fig0S}. Note that the $x$-axis in Fig.S~\ref{Fig0S} is $m_z/m_0$ instead of $m_z$. }
\label{Fig1S}
\end{figure}

In the SCBA, which specializes to weak disorder, the self-energy $\Sigma(E_{F})$ can be obtained by solving the self-consistent integral equation,
\begin{eqnarray}
  \Sigma(E_{F})\! &=&\! \!\frac{W^2 a^2}{96\pi^3}\!  \int_{k \in BZ} d^{3}k [ (E_{F} \!+\! i0^{+}) I \!-\! H_{0} \!-\!\Sigma ]^{-1},
  \label{equ:2}
\end{eqnarray}
where the $k$ integral is over the first Brillouin zone (BZ) and the lattice constant $a\equiv1$. The solution of the self-energy can then be used to obtain the renormalized band dispersion in the SCBA. In solving the integral equation (\ref{equ:2}), we set ${t_{x,y,z}=1,m_0=2.1}$ and $E_{F}=0$ as in the main text. It is straightforward to verify that the Fermi level remains unrenormalized, i.e. $\bar{E}_{F}=0$.

In Fig.S~\ref{Fig1S}, we plot the calculated SCBA band dispersion for different disorder strength $W=\{0,2, 3 ,4\}$ and in systems with different bare mass $m_z=\{-1.2,-1,-0.8,0.8,1,1.2\}$ sitting near the QAH-WSM ($\bar{m}_z/m_0=-t_z/m_0$) and the WSM--NI ($\bar{m}_z/m_0=t_z/m_0$) phase boundaries as indicated by the corresponding roman numerals in Fig.S~\ref{Fig0S}.
Figs.S~\ref{Fig1S}(II-III) show that for $-\bar{m}_z\lesssim t_z$,
the Weyl nodes located at $k^{0}_z=\pm\arccos(\bar{m}_z/t_z)$ are close to the BZ boundary and will move to and annihilate at the BZ boundary ($k_0^z=\pi$) with increasing $W$. As a result, the WSM enters the QAH phase by opening a nontrival (inverted) band gap with $\bar{m}_z<-t_z$ that continues to increase with disorder [see Figs.S~\ref{Fig1S}(I)]. In contrast, for $\bar{m}_z\lesssim t_z$, Figs.S~\ref{Fig1S}(IV-V) show that the Weyl nodes repel each other near the BZ center ($k^{0}_{z}=0$) as the disorder effectively increases their separation by increasing the renormalized mass $\bar{m}_z$. Interestingly, in Fig.S~\ref{Fig1S}(VI) where $\bar{m}_z\gtrsim t_z$, a pair of Weyl nodes is nucleated at zone center with increasing disorder strength, which gives raise to a normal band insulator to WSM transition.

Furthermore, we can obtain a closed form of $\Sigma$ and gain more analytical insights by neglecting $\Sigma$ on the right side of Eq.\ref{equ:2}, which amounts to the Born approximation. Expanding $H_{0}$ to quadratic order in $k$, we obtain
\begin{eqnarray*}
   \bar{m}_{z} &=& m_{z}-\frac{W^2 a^2}{96\pi^2 } \frac{1}{m_{0}} \int_{-\pi}^{\pi} ln\left|(\frac{m_{0}}{t_{z}-t_{z}k_{z}^2-m_{z}})^2
  (\frac{\pi}{a})^4\right| dk_{z}.
\end{eqnarray*}
This result shows that for the parameter regime studied in the main text, the disorder-induced correction to $\bar{m}_{z}$  always has the same sign as $-m_{0}$. The band gap therefore becomes more inverted with increasing disorder strength $W$. In particular, for a WSM with $-t_z<\bar{m}_z<t_z$, the pair of Weyl nodes move to the BZ boundary where $\bar{m}_z$ approaches $-t_z$ and annihilate.
The WSM will then open up an inverted topogolical gap (with $\bar{m}_z<-t_z$) that further increases with the disorder strength $W$. If the clean system is a normal band insulator with $\bar{m}_z>t_z$, the normal band gap will close (when $\bar{m}_z=t_z$) with increasing $W$ by nucleating a pair of Weyl nodes at the zone center that repel each other upon further increasing $W$.